\documentclass{llncs}
\usepackage[english]{babel}
\usepackage[utf8x]{inputenc}
\usepackage[T1]{fontenc}
\usepackage{etoolbox}
\usepackage{amsmath}
\usepackage{graphicx}
\usepackage[colorinlistoftodos]{todonotes}
\usepackage[colorlinks=true, allcolors=blue]{hyperref}

\begin{document}

\title{Peer review and citation data in predicting university rankings, a large-scale analysis}

%\title{Peer Review or Bibliometrics - Which is more effective at predicting institutional rankings?}
%
\titlerunning{Incidental or influential?}  % abbreviated title (for running head)
\author{David Pride \and Petr Knoth
}
\authorrunning{Pride and Knoth.} % abbreviated author list (for running head)
\tocauthor{}
\institute{The Knowledge Media Institute, The Open University, Milton Keynes, UK.\\
\email{\{david.pride, petr.knoth\}@open.ac.uk}
\
\\
}

\maketitle

\begin{abstract}

Most Performance-based Research Funding Systems (PRFS) draw on peer review and bibliometric indicators, two different methodologies which are sometimes combined. A common argument against the use of indicators in such research evaluation exercises is their low correlation at the article level with peer review judgments. In this study, we analyse 191,000 papers from 154 higher education institutes which were peer reviewed in a national research evaluation exercise. We combine these data with 6.95 million citations to the original papers. We show that when citation-based indicators are applied at the institutional or departmental level, rather than at the level of individual papers, surprisingly large correlations with peer review judgments can be observed, up to \textit{r <= 0.802, n = 37, p < 0.001} for some disciplines. In our evaluation of ranking prediction performance based on citation data, we show we can reduce the mean rank prediction error by 25\% compared to previous work. This suggests that citation-based indicators are sufficiently aligned with peer review results at the institutional level to be used to lessen the overall burden of peer review on national evaluation exercises leading to considerable cost savings.

\end{abstract}

\section{Introduction}
%Two types of PRFS
Since the late 20th century there has been a seismic shift in many countries in how research is funded.  In addition to traditional grant or patronage funding, there is growing use of Performance-based Research Funding Systems (PRFS) in many countries. These systems fall largely into two categories; those that focus on peer review judgments for evaluation and those that use a bibliometric approach. The UK and New Zealand both have systems heavily weighted towards peer review. Northern European countries other than the UK tend to favour bibliometric methodologies whereas Italy and Spain consider both peer review judgments and bibliometrics. Research Evaluation Systems overall have dual and potentially dichotomous ends, firstly identifying the best quality research but also, in many cases, the distribution of research funds. There is, however, a large variance in the level of institutional funding granted based on the results of these exercises. The UK's Research Councils distribute £1.6 billion annually entirely on the basis of the results of the Research Excellence Framework (REF) which is the largest single component of university funding.  At the other end of the scale, the distribution of funds based on the results of the Finnish PRFS is just 3\% of the total research budget. Furthermore, the PRFS in Norway and Australia are both used for research evaluation but are not used for funding distribution \cite{hicks2012performance}. Peer-review based PRFS are hugely time-consuming and costly to conduct. In this investigation we ask how well do the results of peer-review based PRFS correlate with bibliometric indicators at the institutional or disciplinary level. A strong correlation would indicate that metrics, where available, can lessen the burden of peer review on national PRFS leading to considerable cost savings, while a weak correlation would suggest each methodology provides different insights.

To our knowledge, this is the first large-scale study exploring the relationship between peer-review judgments and citation data at the institutional level. Our study is based on a new dataset compiled from 190,628 academic papers in 36 disciplines submitted to UK REF 2014, article level bibliometric indicators (6.95m citations) and institutional / discipline level peer-review judgments. This study demonstrates that there is a surprisingly strong correlation between an institutions' \textit{Grade Point Average} (GPA) ranking for its outputs submitted to the UK Research Excellence Framework for many Units of Assessment (UoAs) and citation data. We also shows that this makes it possible to predict institutional rankings with a degree of accuracy in highly cited disciplines.

\section{Related work}
% * <petrknoth@gmail.com> 2018-03-26T10:13:34.040Z:
% 
% > Related work
% We need something here that is looking in these problems but not on REF, i.e. outside of the UK.  DP
% 
% ^ <david.pride@open.ac.uk> 2018-03-28T14:27:03.371Z.
There has long been wide ranging and often contentious discussion regarding the efficacy of both peer review and bibliometrics and whether one or other, or both should be used for Research Evaluation. 
Several other studies have specifically investigated the correlation between the results of different nations' peer review focused Performance-based Research Funding Systems and bibliometric indicators. Anderson \cite{anderson2013evaluating} finds only weak to moderate correlation with results from the New Zealand PRFS and a range of traditional journal rankings. The highest correlation is \textit{r = 0.48} with the Thomson Reuters Journal Citation Report. However Anderson states that this may be due to the much broader scope of research considered by PRFS processes and the additional quality-related information available to panels. Contrary to Anderson, Smith \cite{smith2008benchmarking} used citations from Google Scholar (GS) and correlated these against the results from the New Zealand PRFS in 2008. He found strong correlation, \textit{r = 0.85} for overall PRFS results against Google Scholar citation count. 

A comprehensive global PRFS analysis was conducted by Hicks in 2012. Hicks states there is convincing evidence that when PRFS are used to define league tables this creates powerful incentives for institutions to attempt to 'game' the process, whether in regards to submission selection or staff retention and recruitment policies \cite{hicks2012performance}. A UK government funded report, The Metric Tide, was published in 2015 and gave a range of recommendations for the use of metrics in research evaluation exercises. The Metric Tide study had access to the anonymised scores for the individual submissions to the REF and was therefore directly able to compare on a paper by paper basis the accuracy of a range of bibliometric indicators. This study tested correlations with a range of different bibliometric measures and found correlation with rankings for REF 4* and 3* outputs for some UoAs. Metrics found to have moderately strong correlations with REF scores for a wide range of UoAs included: number of tweets; number of Google Scholar citations; source normalised impact per paper; SCImago journal rank and citation count \cite{m_tide}.

However, The Metric Tide study used different citation metrics and citation data sources from our approach. It is at the institutional UoA level that our study reveals some of the strongest correlations, higher than previously shown. In a related study, Mryglod et al. \cite{Mryglod2015} used departmental h-index aggregation to predict REF rankings. Their work was completed before December 2014 when the REF results were published and contained ranking predictions based on their model with some degree of success. They also experimented by normalising the h-index for each year between 2008 and 2014 but surprisingly found little evidence that timescale played a part in the strength of the correlations they found. An ad hoc study by Bishop \cite{DVB} also found a moderate to strong correlation between departmental research funding based on the results of the UK's Research Assessment and Evaluation (RAE) exercise conducted in 2008, and departmental h-index. Mingers \cite{mingers2017using} recently completed an investigation that collected total citation counts from Google Scholar (GS) for the top 50 academics\footnote{If there were not 50 academics then the total number of academics on GS for that institute was used.} from each UK institute and he found strong correlations with overall REF rankings. To our knowledge, ours is the first large-scale in-depth study that investigates the correlation between citation data and peer review rankings by discipline at the institutional level, taking into account all papers submitted to REF. 

\section{Results}

% How we got the data and how we pre-processed it. 
% Justify why we use MAG

%\subsubsection{Research Excellence Framework - Submissions and Results data}

%In total about 3 paragraphs: Why we used the data, How we acquired it and How the data looks like

%Justify why we used the REF data
% - it is an example of well established system which uses both peer-review and bibliometrics enabling us to compare and contrast these. 
% - lots of data
% - submission data and results data are available (although not at the article level)

% A section of how we created the data (bring workflow Figure 1 as much to the start of the section as possible)

For this study we used data from the UK's Research Excellence Framework (REF). The last REF exercise undertaken in the UK in 2014 was the largest overall assessment of universities' research output ever undertaken globally. These experiments focus on the academic outputs (research papers) component of the REF, for which the metadata are available for download from the REF website. The REF 2014 exercise peer reviewed and graded approximately 191,000 outputs from 154 institutions and in 36 Units of Assessment (UoAs) from zero to four stars. The grading for each submission was determined according to \textit{originality, significance and rigour}. The peer review grades for the individual submissions were aggregated for each UoA to produce a \textit{Grade Point Average} for each institute. The rankings are of critical importance to the institutes as approximately £1.6 billion in QR funding from central government is distributed annually entirely on the basis of the REF results \cite{hefce_panel_a}. 

%, with only 3* and 4* outputs qualifying for inclusion in the funding calculation . 
  
%154 UK Higher Education Institutions made a three-part submission to the REF. These parts individually considered the \textit{outputs} - the actual academic output of the institution, the \textit{impact} of the research that institution undertook, and the \textit{environment} which looked at the places and the practices of research. 

% The REF process convened a panel of expert peer reviewers for each UOA whose role it was to conduct peer review, to international standards, for each submission. In depth detail regarding this process in given in the discussion section.
  
%\subsubsection{Use of citation data in REF peer review process}

Each of the REF peer review panels individually chose whether or not to use citation data to inform their decisions. Eleven out of 36 selected to do so and were provided with citation data from Elsevier Scopus to assist their decision making. For each area and age of publication they were given the number of citations required to put the paper in the top 1\%, 5\%, 10\% or 25\% of papers within its area. Additionally, each journal in the Scopus database is assigned to one or more subject classifications, using their 'All Science Journal Classification' (ASJC) codes. Panels were told the mean number of times that journal articles and conference proceedings published worldwide in that year, in that ASJC code, were cited. This gave REF reviewers a subject-level benchmark against which to consider the citation data.\cite{REF} 

% Additionally, the REF providers noted that;

% \textit{“Citation counts depend partly on the field of research and a publication’s age. Therefore, where sub-panels are making use of citation data the REF team will provide contextual data to assist in the sub-panel’s interpretation of the citation counts. This will consist of information about the citation behaviour of groups of papers worldwide, published in a similar subject area and of a similar age.”} \cite{REF_context}

%\subsubsection{Destruction of individual rankings for outputs}

Whereas the aggregate GPA ranking for all UoAs and all institutes is publicly available, it is now not possible to obtain a direct comparison between citation data and the individual rankings for each submission as HEFCE state that these data were destroyed. The rationale behind this was to preempt any requests for this data under the Freedom of Information Act. \cite{hefce_outputs}.

%This was at least in part to protect individual academics from the potential ramifications if identified as having presented low scoring outputs to the REF \cite{hefce_outputs}.
% * <petrknoth@gmail.com> 2018-04-11T16:34:30.613Z:
% 
% > This was at least in part to protect individual academics from the potential ramifications if identified as having presented low scoring outputs to the REF 
% I think that this sentence can be removed. It is interesting, but it does not add much essential information to the story. 
% 
% ^.
% * <petrknoth@gmail.com> 2018-04-11T16:33:44.781Z:
% 
% > destroyed
% Citation needed!!!
% 
% ^.

% Once all submissions for a Unit of Assessment (UOA) had been graded by the peer review panel, the percentage of unclassified and 0* to 4* submissions for each institute was recorded. A \textit{Grade Point Average} was assigned to each HEI (Higher Education Institute) This was calculated as the mean ranking for each institute for each UOA and was the final figure used to calculate the overall ranking for an institutes' outputs (submissions).
%\subsubsection{REF 2014 Outputs Dataset}

\subsection{Dataset}
% * <petrknoth@gmail.com> 2018-04-11T17:06:01.261Z:
% 
% > Dataset
% I personally think that this section can and should be slightly shortened. 
% 
% ^.

\begin{figure}[t]
\centering
\includegraphics[width=0.9\textwidth]{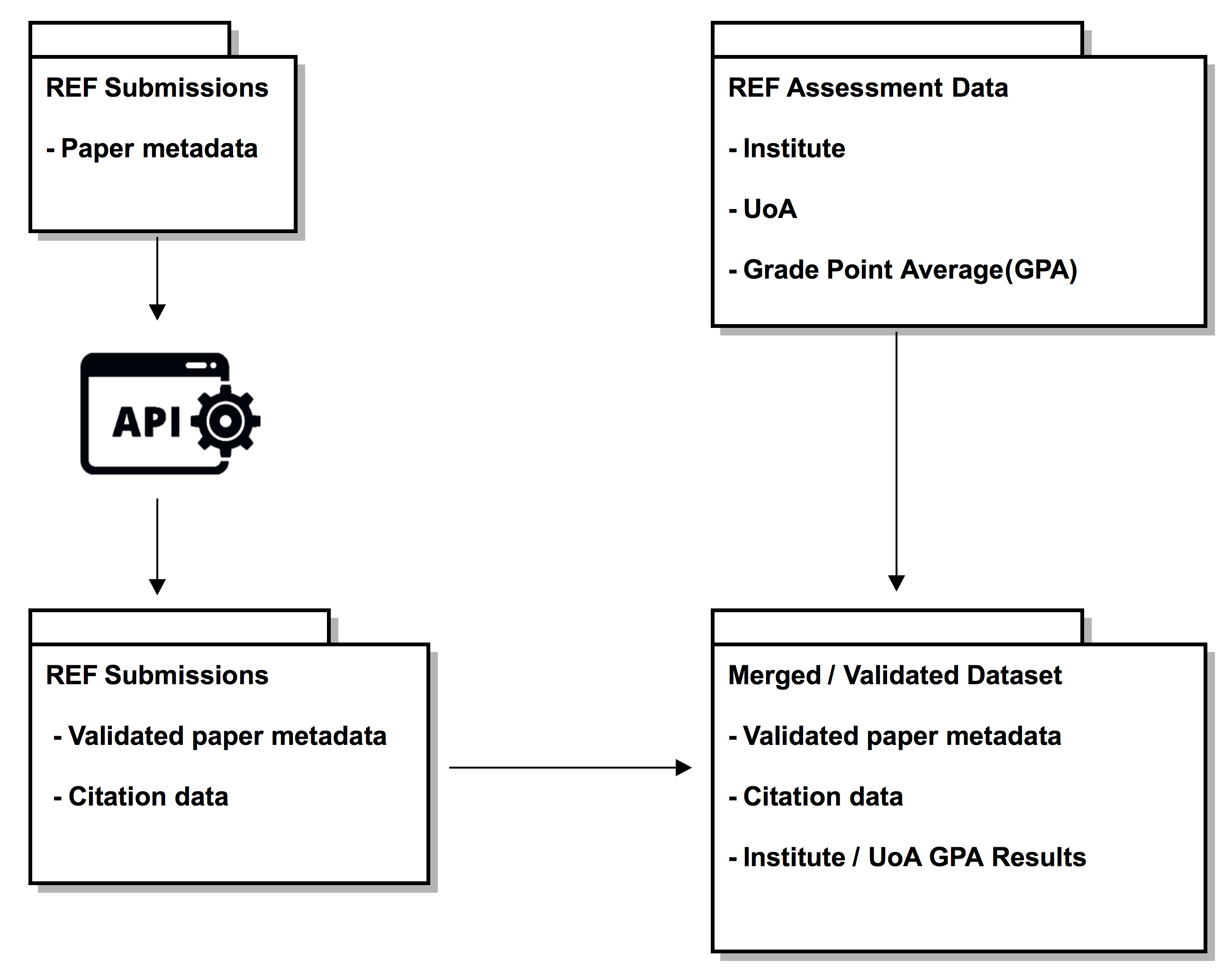}
\caption{\label{workflow} Citation enrichment workflow used in dataset creation.}
\end{figure}

The dataset creation procedure is depicted in Figure \ref{workflow}. We first downloaded the REF 2014 submission list \cite{REF}. For each output, the list contains; publication title, publication year, publication venue, name of institute and UoA. These fields were fully populated for 190,628 out of 190,963 submissions to the 'outputs' category of the REF process. 

%\subsubsection{Collation of citation data - Microsoft Academic Graph}

%In order to test the hypothesis in regards to REF rankings and citation based metrics at the discipline level, it was necessary to collect historic and contemporary citation data for all papers contained within the REF dataset. 
We decided to utilise the Microsoft Academic Graph (MAG) to enrich the REF submission list with citation information. At the time of the experiment MAG contained approximately 168m individual papers and 1.15 billion citation pairs. This decision was motivated by the fact that while Scopus, operated by Elsevier, was used to provide citation data to the REF process, the free version of the Scopus API service is limited to 20,000 requests per week. It would have therefore taken almost two months to gather the required data which was not practical as this was more than 10 times slower than using MAG. Additionally, studies by \cite{herrmannova2016analysis} and \cite{hug2017coverage} have recently confirmed how comprehensive the MAG citation data are. We could not utilise Google Scholar as it does not offer an API and prohibits 'scraping' of data. 

We systematically queried the MAG Evaluate API for each submission using a normalised version of the publication's title (lower case, diacritics removed). This returned a set of MAG IDs which which were potential matches of the article. We subsequently queried the MAG Graph Search API to validate each of the potential matches. We accepted as a match the most similar publication title that had at least $0.9$ cosine similarity. This threshold was set by manually observing about one hundred matches. %TODO: ideally add something like "observing an error rate of less than X \%" 
Using this process we successfully matched 145,415 REF submissions with 6.95 million citations, corresponding to a recall of 76\% of the total initial REF submission list.

\begin{table}[t]
\centering
\begin{tabular}{|l|l|l|l|l|l|}
\hline
\textbf{UoA / Subject}              & \textbf{Outputs} & \textbf{\%  in MAG} & \textbf{Citations} & \textbf{MCPP} \\ \hline
Public Health                       & 4,881                 & 94.61\%             & 505,950             & 109.56      \\ \hline
Clinical Medicine                   & 13,394                & 90.78\%             & 1,278,810            & 105.17    \\ \hline
Physics                             & 6,446                 & 84.51\%             & 491,151             & 90.15       \\ \hline
Biological Sciences                 & 8,608                 & 92.20\%             & 620,009             & 78.12       \\ \hline
Earth Systems / Environment    & 5,249                 & 91.64\%             & 315,429             & 65.58            \\ \hline
Chemistry                           & 4,698                 & 87.71\%             & 246,361             & 59.78        \\ \hline
Allied Health Professions           & 10,358                & 89.35\%             & 402,033             & 43.43          \\ \hline
Ag. Vet. and Food Science           & 3,919                 & 90.76\%             & 150,959             & 42.44             \\ \hline
Comp. Science and Informatics       & 7,645                 & 89.22\%             & 284,815             & 41.76            \\ \hline
Economics and Econometrics          & 2,600                 & 88.81\%             & 95,591              & 41.4              \\ \hline
\end{tabular}
\caption{UoAs with the highest mean citations per paper (MCPP).}
\label{REF_MAG_1}

\begin{tabular}{|l|r|}
\hline
Number of Units of Assessment (UoAs) & 36        \\ \hline
Number of institutes          & 154       \\ \hline
Number of UoAs/institution pairs & 1,911 \\ \hline
Number of submissions (papers)        & 190,628   \\ \hline
Number of submissions (papers) in MAG  & 145,415   \\ \hline
Number of citations           & 6,959,629 \\ \hline
\end{tabular}
\caption{Dataset statistics}
\label{data_stats}
\vspace*{-\baselineskip}
\end{table}

Table \ref{REF_MAG_1} is ordered by the mean citations per paper (MCPP) and shows total number of submissions, percentage of these submissions available in MAG and the total citations of these submissions.

Additionally, as described in Figure \ref{workflow}, we downloaded the Assessment Data from the REF 2014 website. These data contain the GPA, calculated by aggregating the peer review assessment results of individual papers for each given institution per UoA. We then joined these data with the enriched REF submission list by institution name and UoA. By doing so, we obtained 1,911 UoA/institution pairs together with their peer assessment information (GPA) and corresponding lists of submissions and their citation data (Table \ref{data_stats}). 

The full dataset used in our experiments and all results can be downloaded from Figshare. \footnote{https://figshare.com/s/69199811238dcb4ca987}

\subsection{How well do peer review judgments correlate with citation data at the institutional level?} 

% \begin{itemize}
%  \item How well do peer review judgments correlate with citation data at the institutional level for a particular discipline? 

%  %Are peer-review judgments influenced when citation data is available to the reviewers?  
%  \item How well can citation data predict peer review based institutional rankings? %Answer: with a fairly high degree of accuracy
% \end{itemize}

Once we assembled the full dataset, we extracted the following overall citation statistics: mean citations in December 2017 ($mn_{2017}$), median citations in December 2017 ($med_{2017}$), mean citations at the time of the REF exercise ($mn_{2014}$), and median citations at the same point ($med_{2014}$). These data were then used to test the correlation between citation data and REF GPA rankings for outputs for every institute in every UOA. The top ten measured correlations by UoA are shown in Table \ref{correl_table}. The Citation Data (CD) column denotes whether the REF judging panels considered citation data in their deliberations. While we attempted to run correlations with other similar aggregate functions, these are not shown in this table as they have far lower correlations with GPA.

\begin{table}[t]
\centering
\begin{tabular}{|l|l|l|l|l|l|}
\hline
\textbf{CD} & \textbf{UoA}                              & $mn_{2017}$ & $med_{2017}$ & $mn_{2014}$ & $med_{2014}$ \\ \hline
Y            & Chemistry                                 & 0.663                     & \textbf{0.802}     & 0.637                     & 0.738                       \\ \hline
Y            & Biological Sciences                       & 0.188                     & \textbf{0.797}     & 0.288                     & 0.785                       \\ \hline
N            & Aero. Mech. Chem. Engineering             & \textbf{0.771}            & 0.758              & 0.745                     & 0.760                       \\ \hline
N            & Social Work and Policy                    & 0.697                     & \textbf{0.752}     & 0.629                     & 0.635                       \\ \hline
Y            & Comp. Sci. and Informatics                & 0.715                     & \textbf{0.743}     & 0.720                     & 0.678                       \\ \hline
Y            & Economics                                 & \textbf{0.750}            & 0.737              & 0.760                     & 0.770                       \\ \hline
Y            & Earth Systems and Enviro. Sciences        & 0.472                     & \textbf{0.707}     & 0.512                     & 0.686                       \\ \hline
Y            & Clinical Medicine                         & 0.654                     & \textbf{0.677}     & 0.666                     & 0.662                       \\ \hline
Y            & Public Health and Primary Care            & 0.535                     & \textbf{0.674}     & 0.607                     & 0.653                       \\ \hline
Y            & Physics                                   & 0.600                     & \textbf{0.666}     & 0.627                     & 0.605                       \\ \hline
\end{tabular}
\caption{Correlation between REF GPA output rankings and citation data}
\label{correl_table}
\vspace*{-\baselineskip}
\end{table}

\begin{figure}[ht]
\centering
\includegraphics[width=0.8\textwidth]{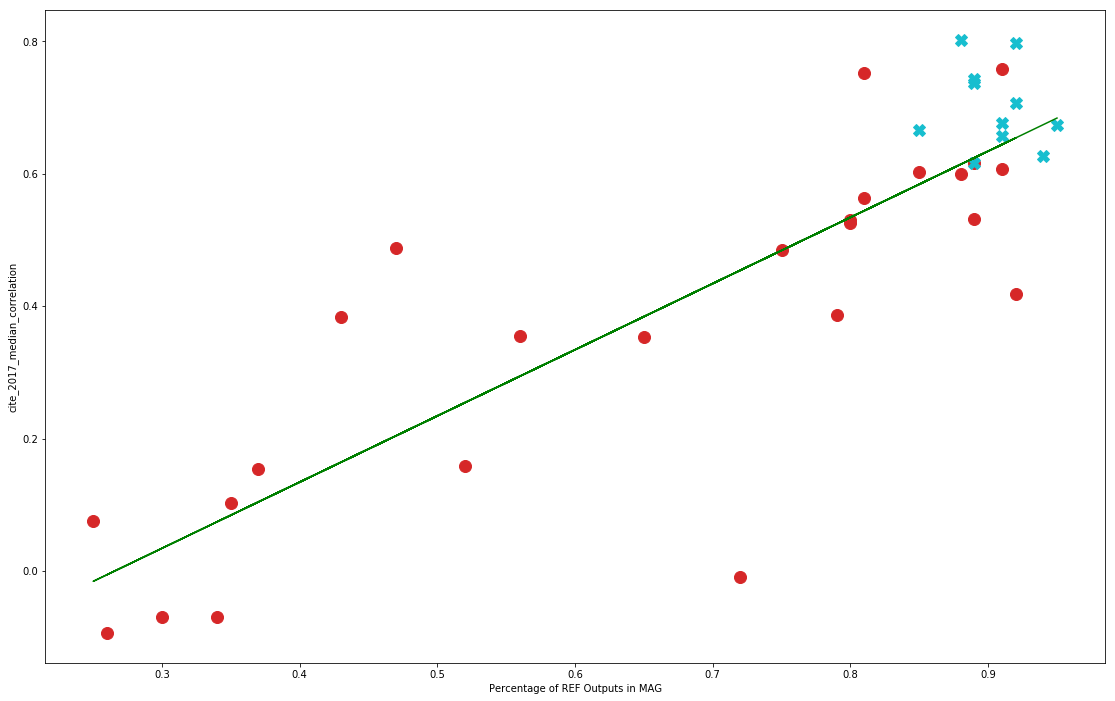}
\caption{\label{fig:1}Correlation between $med_{2017}$ citations per UoA and GPA against the coverage of REF submissions in MAG for all UoAs. An 'o' represents a non-citation based UoA whilst and 'x' denotes a UoA that used citations.}
\label{coverage_plot}
\end{figure}

\begin{table}[ht]
\centering
\begin{tabular}{|l|l|l|l|l|l|l|l|}
\hline
\textbf{REF GPA rank} & \textbf{GPA} & $med_{2017}$ & \textbf{mc2017} & \textbf{rdiff} & $med_{2014}$ & \textbf{mc2014} & \textbf{rdiff} \\ \hline
Liverpool             & 3.44         & Liverpool            & 64              & \textbf{0}                & Liverpool            & 26              & \textbf{0}                \\ \hline
Cambridge             & 3.42         & Cambridge            & 54              & \textbf{0}                & Lancaster            & 25              & +8               \\ \hline
Oxford                & 3.32         & Warwick              & 53              & +3               & Oxford               & 22              & \textbf{0}                \\ \hline
UEA                   & 3.29         & Bath                 & 51              & +12              & Cambridge            & 22              & -2               \\ \hline
Bristol               & 3.26         & Oxford               & 50              & -2               & Queen Mary           & 20              & +2               \\ \hline
\end{tabular}
\caption{Top 5 REF rankings for Chemistry by GPA and predictions produced using $med_{2017}$ and $med_{2014}$ respectively. }
\label{chem_rank}

\centering
\begin{tabular}{|l|l|l|l|l|l|l|l|}
\hline
\textbf{REF GPA Rank} & \textbf{GPA} & $med_{2017}$ & \textbf{mc2017} & \textbf{rdiff} & $med_{2014}$ & \textbf{mc2014} & \textbf{rdiff} \\ \hline
Cambridge             & 3.34         & Cambridge            & 25              & \textbf{0}              & Cambridge           & 9               & \textbf{0}              \\ \hline
Imperial              & 3.12         & Imperial             & 23              & \textbf{0}              & Imperial            & 8               & \textbf{0}              \\ \hline
UCL                   & 3.06         & Sheffield            & 19              & +2             & Brighton            & 7               & +13            \\ \hline
Cranfield             & 3.01         & Brighton             & 18              & +12            & Manchester          & 6               & +4             \\ \hline
Sheffield             & 3.01         & Manchester           & 17              & +3             & Sheffield           & 6               & 0              \\ \hline
\end{tabular}
\caption{Top 5 REF rankings for Aeronautical and Mechanical Engineering by GPA and predictions produced using $med_{2017}$ and $med_{2014}$ respectively.}
\label{aero_rank}
\vspace*{-\baselineskip}
\end{table}

Strong positive correlations can be observed at the discipline level for a large proportion of the UoAs, particularly for median citation count in 2017. Whilst the correlation was most often stronger for those UoAs that had used citation data in the REF peer review process, this was not always the case. Aeronautical and Mechanical engineering and Social work \& Policy are two disciplines, which did not use citation data yet, show very strong correlations with GPA results. 

At the lower end of the scale, there was little correlation between GPA ranking and citation data, notably for those subjects covered by REF panels C and D. The Metric Tide report noted that \textit{There is large variation in the availability of metrics data across the REF submission, with particular issues with coverage in units of assessment (UoAs) in REF Main Panel D} \cite{m_tide}. Lack of coverage in many of these areas is, however, understandable as these are disciplines which do not always produce journal articles, conference proceedings and other digitally published and highly citable artifacts as their main type of output. There is, however, clear delineation between the highly correlated UoAs and those less correlated. The UoAs with the lowest are distinct from the rest, they are having a very weak or no correlation ($r <= 0.159, n = 37, p < 0.001$). Those above this level have a medium to strong correlation ($r > 0.353, n = 37, p < 0.001$).

% Books, stage productions, music, TV and film productions and art installations all fall into this category. Although impactful in their own right, none of these examples can be assessed accurately or fairly with standard bibliometrics.

The variance of citation data coverage across UoAs led us to explore whether there could be a relationship between the strength of the correlations GPA and citation data correlation with the coverage of citation in a given UoA. Figure \ref{coverage_plot} plots this for both the UoAs that used citation data and those that did not. While the graph confirms that the highly cited UoAs in MAG are those UoAs that used citation data, it indicates that a few  UoAs that did not also exhibit strong correlations. Unsurprisingly, the plot suggests that there might be a small bias exhibited by extra correlation strength in UoAs that utilised citation data. However, given the small number of UoAs, this is not statistically significant. 

\begin{table}[ht]
\centering
\begin{tabular}{|l|l|l|l|l|l|l|l|l|l|}
\hline
\textbf{UoA}   & \textbf{HEIs} & \textit{\textbf{rdiff}} & \textit{\textbf{nrdiff}} & \textit{\textbf{MAP}} & \textit{\textbf{MAP}} & \textit{\textbf{MAP}} & \textit{\textbf{MAP}} & \textit{\textbf{MAP}} & \textit{\textbf{MAP}} \\ 
   & &  &  & \textit{\textbf{rt=3}} & \textit{\textbf{rt=5}} & \textit{\textbf{rt=10}} & \textit{\textbf{rt=10\%}} & \textit{\textbf{rt=20\%}} & \textit{\textbf{rt=30\%}} \\ \hline
Comp Sci.      & 89            & 12.39          & 0.139             & 0.19          & 0.32          & 0.50           & 0.46             & 0.75             & 0.87             \\ \hline
Ag. Vet.       & 29            & 4.02           & 0.139             & 0.45          & 0.65          & 0.86           & 0.45             & 0.68             & 0.86             \\ \hline
Clinical Med.  & 31            & 4.38           & 0.141             & 0.51          & 0.70          & 0.93           & 0.51             & 0.77             & 0.93             \\ \hline
Allied H.      & 83            & 12.03          & 0.145             & 0.20          & 0.30          & 0.55           & 0.43             & 0.72             & 0.86             \\ \hline
Economics      & 28            & 4.07           & 0.145             & 0.57          & 0.71          & 0.92           & 0.57             & 0.78             & 0.92             \\ \hline
Chemistry      & 37            & 5.51           & 0.149             & 0.54          & 0.56          & 0.83           & 0.54             & 0.78             & 0.86             \\ \hline
Earth Systems  & 45            & 7.24           & 0.161             & 0.40          & 0.51          & 0.77           & 0.51             & 0.68             & 0.84             \\ \hline
Public Health  & 32            & 5.18           & 0.162             & 0.50          & 0.62          & 0.84           & 0.50             & 0.68             & 0.84             \\ \hline
Bio. Science   & 44            & 7.59           & 0.173             & 0.34          & 0.52          & 0.72           & 0.52             & 0.66             & 0.79             \\ \hline
Physics        & 41            & 7.36           & 0.180             & 0.36          & 0.53          & 0.78           & 0.43             & 0.73             & 0.80             \\ \hline
\textbf{All (mean)} & \textbf{45}   & \textbf{6.98}  & \textbf{0.153}    & \textbf{0.41} & \textbf{0.54} & \textbf{0.77}  & \textbf{0.49}    & \textbf{0.72}    & \textbf{0.86}    \\ \hline
\end{tabular}
\caption{Rank prediction quality for top 10 UoAs with the highest mean citations per paper.}
\label{prediction_quality}
\vspace*{-\baselineskip}
\end{table}

\subsection{How well can citation data predict peer review based institutional rankings?}
Tables \ref{chem_rank} and \ref{aero_rank} shows top 5 institutions for Chemistry and Aeronautical and Mechanical Engineering as ranked in the REF by GPA and predictions of ranking using $med_{2017}$ and $med_{2014}$ respectively. \textit{mc2017} and \textit{mc2014} show the median citation count for that institute. \textit{Rdiff} shows the rank difference  when ranked by a particular citation metric. The prediction performance indicated in these tables is not unique, in four of the five top UoAs by correlation strength the highest ranked institute is predicted correctly by both $med_{2014}$ and $med_{2017}$. 

Table \ref{prediction_quality} demonstrates the effectiveness of predicting based on $med_{2014}$ for the 10 most highly cited UoAs. To compare the prediction error, expressed by \textit{rdiff}, across UoAs, we calculated the mean rank difference normalised by number of institutions (\textit{nrdiff}). To express overall prediction accuracy, we used Mean Average Precision (MAP). The parameter $rt$ denotes the prediction rank tolerance. For example, $rt=3$ indicates that a prediction within 3 positions of the original assessment result will be considered as correct. Given the simplicity of the prediction method, this is a strong indication of the power of citation data in this task. One could reasonably expect that further improvements can be made by employing more sophisticated indicators. However, as the predictions are not as good for UoAs that have lower than average mean citations per paper , we would restrain from recommending the use of citation data unaccompanied by peer review assessments in those UoAs. 

%TODO MArygold
%We reduced our collection to include institutions which Myrgold had in their study. This is because they were unable to obtain as comprehensive data as we collected.  
%We reduced the error (rdiff) reported by Myrgold by, our MAP is higher than Myrgold by ... 
We wanted to compare our prediction performance to the study of Mryglod et al. \cite{Mryglod2015}. In order to conduct a fair and exact comparison, it was necessary to parse a number of institutions from our input data. Mryglod et al. reported they were unable to obtain citation indicators for all institutions in a given UoA. Their study covered three of the top ten highly cited UoAs, we show in Table \ref{comparison} that our predictions are significantly better across all categories. 

%on a set of UoAs with high citations per paper, which is actually those UoAs for which, as our analysis indicates, citation data have the potential to provide good prediction perfomance. 

\begin{table}[t]
\centering
\begin{tabular}{|l|l|l|l|l|l|l|l|l|l|}
\hline
\textbf{UoA}   & \textbf{HEIs} & \textit{\textbf{rdiff}} & \textit{\textbf{nrdiff}} & \textit{\textbf{MAP}} & \textit{\textbf{MAP}} & \textit{\textbf{MAP}} & \textit{\textbf{MAP}} & \textit{\textbf{MAP}} & \textit{\textbf{MAP}} \\ 
   & &  &  & \textit{\textbf{rt=3}} & \textit{\textbf{rt=5}} & \textit{\textbf{rt=10}} & \textit{\textbf{rt=10\%}} & \textit{\textbf{rt=20\%}} & \textit{\textbf{rt=30\%}} \\ \hline
\multicolumn{10}{ |l| }{\textbf{Mryglod} \cite{Mryglod2015}} \\ \hline
Chemistry  & 29            & 4.89           & 0.169             & 0.37          & 0.82          & 0.82           & 0.37             & 0.82             & 0.82             \\ \hline
Physics  & 32            & 8.63           & 0.270             & 0.28          & 0.40          & 0.65           & 0.28             & 0.46             & 0.65             \\ \hline
Bio Science   & 31            & 8.38           & 0.270             & 0.22          & 0.38          & 0.70           & 0.22             & 0.51             & 0.64             \\ \hline
\textbf{All (mean)} & \textbf{31}   & \textbf{7.30}  & \textbf{0.24}     & \textbf{0.29} & \textbf{0.53} & \textbf{0.72}  & \textbf{0.29}    & \textbf{0.60}    & \textbf{0.70}    \\ \hline
\multicolumn{10}{ |l| }{\textbf{Pride \& Knoth (this study)}} \\ \hline
Chemistry  & 29            & 4.00           & 0.138             & 0.68          & 0.72          & 0.89           & 0.68             & 0.72             & 0.86             \\ \hline
Physics  & 32            & 5.68           & 0.178             & 0.34          & 0.59          & 0.90           & 0.34             & 0.75             & 0.90             \\ \hline
Bio Science   & 31            & 7.16           & 0.231             & 0.35          & 0.45          & 0.74           & 0.35             & 0.51             & 0.71             \\ \hline
\textbf{All (mean)} & \textbf{31}   & \textbf{5.61}  & \textbf{0.18}     & \textbf{0.46} & \textbf{0.59} & \textbf{0.84}  & \textbf{0.46}    & \textbf{0.66}    & \textbf{0.82}    \\ \hline
\textbf{Improvement} &   & \textbf{23\%}  & \textbf{25\%} & \textbf{59\%}	& \textbf{11\%}	& \textbf{17\%} & \textbf{59\%} & \textbf{10\%}	& \textbf{17\%} \\ \hline
\end{tabular}
\caption{Comparison of the prediction performance of our study with Mryglod et al.\cite{Mryglod2015}}
\label{comparison}
\vspace*{-\baselineskip}
\end{table}

\section{Discussion}

It has been shown in \cite{m_tide}, \cite{baccini2016} and that many bibliometric indicators show little correlation with peer review judgments at the article level. This study, and those by \cite{mingers2017using}, \cite{anderson2013evaluating} and \cite{smith2008benchmarking}, demonstrate that some bibliometric measures can offer a surprisingly high degree of accuracy when used at the institutional or departmental level. Our work has been conducted on a significantly larger dataset and our prediction accuracy is higher than shown in previous studies, despite deliberately using fairly simplistic indicators. 

Several studies including The Metric Tide \cite{m_tide}, The Stern Report \cite{stern} and the HEFCE pilot study \cite{hefce_pilot} all state that metrics should be used as an additional component in research evaluation, with peer review remaining as the central pillar. Yet, peer review has been shown by \cite{hojat2003impartial}, \cite{lee2013bias} and \cite{smith2006peer} amongst others to exhibit many forms of bias including institutional bias, gender / age related bias and bias against interdisciplinary research. In an examination of one of the most critical forms of bias, that of publication bias, Emerson \cite{emerson2010testing} noted that reviewers were much more likely to recommend papers demonstrating positive results over those that demonstrated null or negative results.

All of the above biases exist even when peer review is carried out to the highest international standards. There were close to 1,000 peer review experts recruited by the REF, however the sheer volume of outputs requiring review calls into question the exactitude of the whole process. As an example the REF panel for UoA 9, Physics, consisted of 20 members. The total number of outputs submitted for this UoA was 6,446. Each paper is required to be read by two referees. This increases the overall total requirement to read 12,892 paper instances. Therefore each panel member was required to review, to international standards, an average of 644 papers in a little over ten months. If every panel member, worked every day for ten months, each member would need to read and review 2.14 papers \textit{per day} to complete the work on time. This is, of course, in addition to the panelist's usual full-time work load. Moreover, Physics is not an unusual example and many other UoAs tell a similar story in terms of the average number of papers each panel member was expected to review; Business and Management Studies (1,017 papers), General Engineering (868 papers), Clinical Medicine (765 papers). The burden placed on the expert reviewers during the REF process was onerous in the extreme. Coles \cite{coles2013} calculated a very similar figure of 2 papers per day, based on an estimate before the data we now have was available. 'It is blindingly obvious,' he concluded, 'that whatever the panels do will not be a thorough peer review of each paper, equivalent to refereeing it for publication in a journal'. Sayer \cite{sayer2014rank} is equally disparaging in regards to the volume of papers each reviewer was required to read and also expresses significant doubts about the level of expertise within the review panels themselves. 

% This calls out into question the way how peer review judgments were made. 
% We can already see from Figure 2 that panels using citation data tend to have higher correlation than those that didn't. Up to what extent were peer-reviewers influenced by the availability of the citation data? 
% However, the correlations are high even in some panels that did not use these data, although not as high as the panels that did. Could it be the case that despite the fact that peer-reviewers were asked not to use citation data, some did, possibly due to their enormous work load. Or, is it the case that peer reviews really truly correlate so well with citation data at the institutional level? We cannot answer this question as we can only analyse the data of the study in the way it was conducted.  

% This is in clear contrast to the official line.....

% 'It was remarkable to note that panel members had thoroughly read all of the publications assigned to them. It was clear that, despite the large number of papers that each was assigned to review, the evaluators could discuss the manuscripts they had reviewed in depth. \textit{Some had as many as 900 outputs to review}.'(Italics added) \cite{hefce_panel_a}

%\subsection{The cost of Performance-based Research Funding Systems}

In addition to the potential pitfalls in the current methodologies, there is also the enormous cost to be considered. This was estimated to be £66m for the UK's original PRFS, the Research Assessment Exercise (RAE) in 2008. This rose markedly to £246m for the 2014 Research Excellence Framework. This is comprised of £232M in costs to the higher education institutes and around £14M in costs for the four UK higher education funding bodies. The cost to the institutions was approximately £212M for preparing the REF submissions for the three areas; outputs, impact and environment, with the cost for preparing the outputs being the majority share of this amount. Additionally, there were costs of around £19M for panelists’ time. \cite{ref_costs}. If bibliometric indicators can in any way lessen the financial burden of these exercises on the institutions this is a strong argument in favour of their usage. 
\vspace*{-\baselineskip}
%Gaming REF? 

\section{Conclusion}

This work constitutes the largest quantitative analysis of the relationship between peer reviews (190,628 paper submissions) and citation data (6.9m citation pairs) at an institutional level. Firstly, our results show that citation data exhibit strong correlations with peer review judgments when considered at the institutional level and within a given discipline. These correlations tend to be higher in disciplines with high mean citations per paper. Secondly, we demonstrate that we can utilise citation data to predict top ranked institutions with a surprisingly high precision. In the ten UoAs with the highest number of mean citations per paper we achieve 0.77 MAP with prediction rank tolerance 10 with respect to the REF 2014 results. Additionally, in four out of five top UoAs by correlation strength, the highest ranked institute in the REF results was predicted correctly. It is also important to note that these predictions are based on citation data that were available at the time of the REF exercise. 

While our analysis does not answer whether using citation-based indicators we can predict institutional rankings better than by relying on a peer review system, our results evidence that the REF peer review process led to highly similar results as those that could have been predicted automatically using citation data. The 11 REF UoAs with the highest mean citations per paper in MAG are the identical UoAs in which the peer review panels used citation data to inform their decisions. We argue that if peer-review is conducted in the way it was conducted in the REF, then it would have been more cost effective to save a significant proportion of the £246m spent on organising the peer review process  \cite{ref_costs} and carry out the institutional evaluation purely using citation data, particularly in UoAs with high mean citations per paper. 

This has wide implication for PRFS globally. The countries whose PRFS still have a peer review component should carefully consider the way in which the peer review process is conducted. Thus ensuring that the peer review results add a new dimension to the information over that which can be obtained by predictions based on citation data alone. However, this advice only applies when the goal of the PRFS is to rank institutions, as it is the case in the UK REF, rather than individual papers or researchers.

\section{Acknowledgements}
This work has been funded by Jisc and has also received support from the scholarly communications use case of the EU OpenMinTeD project under the H2020-EINFRA-2014-2 call, Project ID: 654021

\bibliographystyle{vancouver}
\bibliography{biblio}

\begin{thebibliography}{10}

\bibitem{hicks2012performance}
Hicks D.
\newblock Performance-based university research funding systems.
\newblock Research policy. 2012;41(2):251--261.

\bibitem{anderson2013evaluating}
Anderson DL, Smart W, Tressler J.
\newblock Evaluating research--peer review team assessment and journal based
  bibliographic measures: New Zealand PBRF research output scores in 2006.
\newblock New Zealand Economic Papers. 2013;47(2):140--157.

\bibitem{smith2008benchmarking}
Smith AG.
\newblock Benchmarking Google Scholar with the New Zealand PBRF research
  assessment exercise.
\newblock Scientometrics. 2008;74(2):309--316.

\bibitem{m_tide}
HEFCE. The Metric Tide: Report of the Independent Review of the Role of Metrics
  in Research Assessment and Management; 2015.
\newblock Available from:
  \url{http://www.hefce.ac.uk/pubs/rereports/year/2015/metrictide/}.

\bibitem{Mryglod2015}
Mryglod O, Kenna R, Holovatch Y, Berche B.
\newblock Predicting results of the Research Excellence Framework using
  departmental h-index.
\newblock Scientometrics. 2015 Mar;102(3):2165--2180.
\newblock Available from: \url{https://doi.org/10.1007/s11192-014-1512-3}.

\bibitem{DVB}
Bishop D. An alternative to REF2014?; 2013.
\newblock Available from:
  \url{http://deevybee.blogspot.co.uk/2013/01/an-alternative-to-ref2014.html}.

\bibitem{mingers2017using}
Mingers J, O’Hanley JR, Okunola M.
\newblock Using Google Scholar institutional level data to evaluate the quality
  of university research.
\newblock Scientometrics. 2017;113(3):1627--1643.

\bibitem{hefce_panel_a}
HEFCE. Research Excellence Framework 2014: Overview report by Main Panel A and
  Sub-panels 1 to 6; 2015.
\newblock Available from:
  \url{http://www.ref.ac.uk/2014/media/ref/content/expanel/member/Main%20Panel%20A%20overview%20report.pdf}.

\bibitem{REF}
HEFCE. Research Excellence Framework - Results and Submissions; 2014.
\newblock Available from: \url{http://results.ref.ac.uk/Results}.

\bibitem{hefce_outputs}
HEFCE. Annex A - Summary of additional information about outputs; 2014.
\newblock Available from:
  \url{http://www.ref.ac.uk/2014/media/ref/content/pub/panelcriteriaandworkingmethods/01_12a.pdf}.

\bibitem{herrmannova2016analysis}
Herrmannova D, Knoth P.
\newblock An analysis of the microsoft academic graph.
\newblock D-Lib Magazine. 2016;22(9/10).

\bibitem{hug2017coverage}
Hug SE, Br{\"a}ndle MP.
\newblock The coverage of Microsoft Academic: Analyzing the publication output
  of a university.
\newblock Scientometrics. 2017;113(3):1551--1571.

\bibitem{baccini2016}
Baccini A, De~Nicolao G.
\newblock Do they agree? Bibliometric evaluation versus informed peer review in
  the Italian research assessment exercise.
\newblock Scientometrics. 2016;108(3):1651--1671.

\bibitem{stern}
Stern N, et~al.
\newblock Building on success and learning from experience: an independent
  review of the Research Excellence Framework.
\newblock London: UK Government, Ministry of Universities and Science. 2016;.

\bibitem{hefce_pilot}
HEFCE. Report on the pilot exercise to develop bibliometric indicators for the
  Research Excellence Framework; 2016.

\bibitem{hojat2003impartial}
Hojat M, Gonnella JS, Caelleigh AS.
\newblock Impartial judgment by the “gatekeepers” of science: fallibility
  and accountability in the peer review process.
\newblock Advances in Health Sciences Education. 2003;8(1):75--96.

\bibitem{lee2013bias}
Lee CJ, Sugimoto CR, Zhang G, Cronin B.
\newblock Bias in peer review.
\newblock Journal of the Association for Information Science and Technology.
  2013;64(1):2--17.

\bibitem{smith2006peer}
Smith R.
\newblock Peer review: a flawed process at the heart of science and journals.
\newblock Journal of the royal society of medicine. 2006;99(4):178--182.

\bibitem{emerson2010testing}
Emerson GB, Warme WJ, Wolf FM, Heckman JD, Brand RA, Leopold SS.
\newblock Testing for the presence of positive-outcome bias in peer review: a
  randomized controlled trial.
\newblock Archives of internal medicine. 2010;170(21):1934--1939.

\bibitem{coles2013}
Coles P. The apparatus of research assessment is driven by the academic
  publishing industry; 2013.
\newblock Available from: \url{https://bit.ly/2EfNMeV}.

\bibitem{sayer2014rank}
Sayer D.
\newblock Rank hypocrisies: The insult of the REF.
\newblock Sage; 2014.

\bibitem{ref_costs}
Technopolis. REF Accountability Review: Costs, benefits and burden; 2015.

\end{thebibliography}

% \begin{table}[]
% \centering
% \caption{Strongest and weakest aggregate correlations for all UOAs.}
% \label{ref_cite_diff}
% \begin{tabular}{|l|r|r|r|}
% \hline
%                                       & \multicolumn{1}{l|}{\textbf{Citation Data}} & \multicolumn{1}{l|}{\textbf{No Citation Data}} & \multicolumn{1}{l|}{\textbf{Difference}} \\ \hline
% \textbf{Most conservative estimate:}  & 0.799                                                 & 0.755                                                    & 0.044                                    \\ \hline
% \textbf{Least conservative estimate:} & 0.700                                                 & 0.368                                                    & 0.332                                    \\ \hline
%\end{tabular}
%\end{table}

\end{document}